\documentclass[pra,aps,showpacs,twocolumn,floats,floatfix,superscriptaddress]{revtex4}
\usepackage{amsfonts}
\usepackage{amsmath}
\usepackage{amssymb}
\usepackage{graphicx}
\usepackage{txfonts}

\begin{document}

\title{Quantum-enhanced microscopy with binary-outcome photon counting}
\author{G. R. Jin}
\email{grjin@bjtu.edu.cn}
\affiliation{Department of Physics, Beijing Jiaotong University, Beijing 100044, China}
\author{W. Yang}
\email{wenyang@csrc.ac.cn}
\affiliation{Beijing Computational Science Research Center, Beijing 100084, China}
\author{C. P. Sun}
\email{cpsun@csrc.ac.cn}
\affiliation{Beijing Computational Science Research Center, Beijing 100084, China}
\date{\today }

\begin{abstract}
Polarized light microscopy using path-entangled $N$-photon states (i.e., the
N00N states) has been demonstrated to surpass the shot-noise limit at very
low light illumination. However, the microscopy images suffer from
divergence of phase sensitivity, which inevitably reduces the image quality.
Here, we show that due to experimental imperfections, such a singularity
also takes place in the microscopy that uses twin-Fock states of light for
illumination. We propose two schemes to completely eliminate this
singularity: (i) locking the phase shift sensed by the beams at the optimal
working point, by using a spatially dependent offset phase; (ii) a
combination of two binary-outcome photon counting measurements, one with a
fixed offset phase and the other without any offset phase. Our observations
remain valid for any kind of binary-outcome measurement and may open the way
for quantum-enhanced microscopy with high $N$ photon states.
\end{abstract}

\pacs{42.50.Dv, 06.30.Bp, 42.50.St}
\maketitle

\section{Introduction}

Light microscopy at low light illumination is desirable to avoid damaging
the specimen (e.g., the biological samples)~\cite%
{Brida,Taylor,Lemos,Ono,Israel}. At very low light level, it might be more
efficient to use nonclassical light for illumination, such as twin beams
from a parametric down-converted light~\cite{Brida} and amplitude squeezed
light~\cite{Taylor}. Recently, polarized light microscopy using
path-entangled $N$-photon states (i.e., the N00N states) $\sim (|N,0\rangle
+|0,N\rangle )$ was demonstrated to enlarge the contribution of each photon
to the image contrast~\cite{Ono,Israel}, where $|m,n\rangle \equiv |m\rangle
_{H}\otimes |n\rangle _{V}$ denotes the product of photon Fock states of two
orthogonal polarization modes $H$ and $V$. From binary-outcome photon
counting~\cite{Ono,Israel}, it was found that the birefringence phase shift
of a sample $\phi (x,y)$ can be estimated beyond the shot-noise limit, i.e.,
the phase sensitivity $\delta \phi (x,y)<1/\sqrt{N}$. However, the phase
sensitivity diverges at certain values of phase shift, which in turn reduces
the quality of microscopy images~\cite{Israel}.

Compared with the N00N states, the twin-Fock states $|n,n\rangle $ are
easier to prepare and more robust against photon loss~\cite%
{HB,FWSun,GYXiang10,Xiang}. Recently, it was shown that the visibility of
the 6-photon count rate could reach $\sim 94\%$~\cite{Xiang}, significantly
better than that of a five-photon N00N state~\cite{Afek}. In addition, the
achievable phase sensitivity can surpass that of the N00N states with a
binary-outcome photon counting~\cite{Xiang}. Similar to Ref.~\cite{Israel},
however, we will show that quantum-enhanced microscopy illuminated by the
twin-Fock state of the light (or any finite-$N$ input state) also suffers
from the divergence of the phase sensitivity. To remedy this problem, we
propose a scheme to lock the phase shift sensed by the beams at the optimal
working point using three estimators nearby, as illustrated schematically by
Fig.~\ref{fig1}(a). We further show that a combination of two binary-outcome
photon counting, one with a fixed offset phase and the other without any
offset phase, also works to remove the singularity. Our results can be
generalized to any kind of binary-outcome measurement that has been widely
adopted in quantum metrology~\cite%
{Bollinger,Dowling,Gerry,Cohen,Brivio,Distante,XMFeng}, and recently in
quantum-enhanced microscopy~\cite{Ono,Israel}.

\section{Binary-outcome photon counting using twin-Fock states of light}

\begin{figure}[tbph]
\begin{centering}
\includegraphics[width=1\columnwidth]{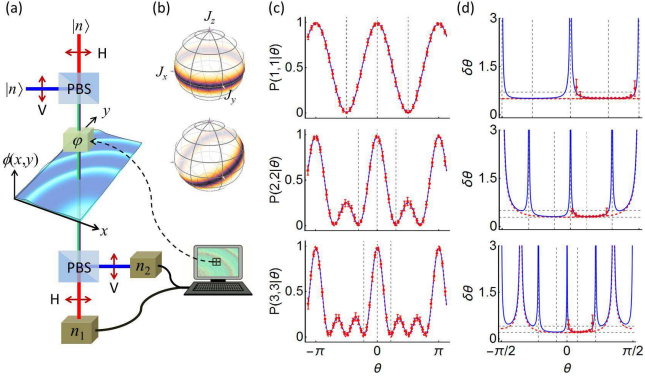}
\caption{(a) Polarized light microscopy with a feedback offset phase. (b)
Quasi-probability distributions of the input $|n,n\rangle $ and the output $%
\exp (-i\protect\theta J_{y})|n,n\rangle $ on the Poincar\'{e} sphere, where
$\protect\theta =\protect\varphi +\protect\phi (x,y)$. (c) Statistical average of the count rate (red circles) and its standard deviation (bars) from $\mathcal{N}=100$ measurements and $20$ repetitions. (d) Phase uncertainty of the maximum likelihood estimator (red circles) and the phase sensitivity (blue solid) using $P_{\mathrm{fit}}(n,n|\protect\theta)$. The red dashed line: the sensitivity with the exact $P(n,n|\theta)$. Horizontal grid lines: shot-noise limit $1/\protect\sqrt{N}$ and $\protect%
\delta \protect\theta _{\mathrm{QCRB}}$ for $N=2n=2$, $4$, $6$.}
\label{fig1}
\end{centering}
\end{figure}

As illustrated schematically in Fig.~%
\ref{fig1}(a), we consider a quantum-enhanced microscopy illuminated by the twin-Fock states of light $|n,n\rangle $~\cite{HB,FWSun,GYXiang10,Xiang}, with the number of photons $N=2n$. The microscopy
images can be constructed from the coincidence photon counting at the output
ports~\cite{Ono,Israel}. Theoretically, the conditional probability for detecting $n_{1}$ photons in
the $H$ polarization mode and $n_{2}$ photons in the $V$ polarization mode
is given by
\begin{equation}
P(n_{1},n_{2}|\theta )=\left\vert \langle n_{1},n_{2}|e^{-i[\varphi +\phi
(x,y)]J_{y}}|n,n\rangle \right\vert ^{2},  \label{Pn1n2}
\end{equation}%
where $\varphi$ is a controllable offset phase, $\phi (x,y)$ is the spatially
dependent phase shift caused by the birefringence of the polarized beams
inside the sample~\cite{Israel}, and $\theta (x,y)\equiv \varphi +\phi (x,y)$%
. The phase accumulation $\exp (-i\theta J_{y})$ can be implemented with a polarization Mach-Zehnder interferometer~\cite{Yurke,Sanders95,Kim98}, corresponding to a
rotation around the $y$-component of the Stokes vector $\boldsymbol{J}%
=(a_{H}^{\dagger },a_{V}^{\dagger })\boldsymbol{\sigma }(a_{H},a_{V})^{T}/2$%
, where $a_{H}$ ($a_{V}$) is the annihilation operator of the polarization
mode $H$ ($V$), and $\boldsymbol{\sigma }$ denotes the Pauli operator.

The photon detection event $n_{1}=n_{2}=n$ is of interest~\cite%
{FWSun,GYXiang10,Xiang} and is denoted as the outcome \textquotedblleft $+$%
\textquotedblright . This is indeed a projection measurement, or
equivalently, a binary-outcome measurement (see Appendix A). The output
signal is $\langle \mu (\theta )\rangle \equiv \langle \psi (\theta )|\mu
|\psi (\theta )\rangle =P(n,n|\theta )$, where $\mu =|n,n\rangle \langle n,n|
$ and $|\psi (\theta )\rangle =\exp (-i\theta J_{y})|\psi _{\mathrm{in}%
}\rangle $. For each given phase shift $\theta \in (-\pi ,\pi )$, after $%
\mathcal{N}$ binary-outcome measurements, the signal is measured by the
count rate $P(n,n|\theta )\simeq \mathcal{N}_{+}/\mathcal{N}$, where $%
\mathcal{N}_{+}$ is the occurrence number of the event $n_{1}=n_{2}=n$. In
Fig.~\ref{fig1}(c), we show the statistical average of $\mathcal{N}_{+}/%
\mathcal{N}$ and its standard deviation (the circles and the bars) obtained
from numerical simulation: first we generate $\mathcal{N}$ random numbers $%
\{\xi _{1},\xi _{2},...,\xi _{\mathcal{N}}\}$~\cite{Pezze&Smerzi} uniformly
distributed within $[0,1]$, then we obtain the occurrence number $\mathcal{N}%
_{+}$ as the number of counts for $\xi $ to lie within the interval $%
[0,P(n,n|\theta )]$. Here, to take the experimental imperfections into
account, we have replaced $P(n,n|\theta )$ with $a_{0}P(n,n|\theta )+b_{0}$,
with $a_{0}$ and $b_{0}$ related to the imperfect visibility and reduced
peak height at the phase origin, respectively (see Appendix B). As depicted
in Fig.~\ref{fig1}(c), the averaged signal, fitted by $P_{\mathrm{fit}%
}(n,n|\theta )$, show multifold oscillations and the first dark fringe
appears at $\theta _{\mathrm{dark}}\simeq \pi /2$, $\arccos (\sqrt{1/3})$,
and $\arctan (\sqrt{2/3})$, from the top to the bottom.

The microscopy images can be reconstructed from the inversion phase estimator~\cite{Israel}, which is a solution of the equation $
P(n,n|\theta )=\mathcal{N}_{+}/\mathcal{N}$ (see Appendix A). To avoid the phase ambiguity~%
\cite{Pezze,Higgins,Berry}, we assume that true value of the phase shift lies within a
monotonic regime of $P(n,n|\theta )$, e.g., $\theta \in (0,\theta _{\mathrm{dark}})$. The image quality is determined
by the phase uncertainty $\delta \theta =\Delta \mu /|\partial \langle
\mu (\theta )\rangle /\partial \theta |=1/\sqrt{F(\theta )}$, where, for a
single-shot measurement, the fluctuations of signal $(\Delta \mu )^{2}\equiv
\langle \mu ^{2}\rangle -\langle \mu \rangle ^{2}=P(n,n|\theta
)[1-P(n,n|\theta )]$ and $F(\theta )$ is the classical Fisher information of
the binary-outcome photon counting measurements (see Appendix A). In Fig.~%
\ref{fig1}(d), we plot the phase sensitivity as a function of $\theta $,
using the exact (fitted) expression of $P(n,n|\theta )$. For the exact cases
(the red dashed lines), the sensitivity reaches minimum at $\theta =0$~\cite%
{FWSun}. Due to the experimental imperfections, however, the best
sensitivity occurs at $\theta _{\min }\simeq 0.88$, $0.37$, and $0.26$ ($%
\sim 15^{\circ }$~\cite{Xiang}), from the top to the bottom, as depicted by
the blue solid lines of Fig.~\ref{fig1}(d).

At the optimal working point $\theta _{\min }$, the sensitivity can surpass
the shot-noise limit by an enhancement factor $\eta =1/(\sqrt{N}\delta
\theta _{\min })\simeq 1.39$ (for $N=2$), $1.61$ ($N=4$), and $1.85$ ($N=6$%
). Theoretically, the enhancement factor can be predicted by calculating the
quantum Fisher information of a phase-encoded state $\exp (-i\theta G)|\psi
_{\mathrm{in}}\rangle $~\cite{Helstrom,Braunstein,Giovannetti}, where $G$ is
a hermitian operator that encodes a phase shift on the input state $|\psi _{%
\mathrm{in}}\rangle $. The optimal choice of $G$ is fully determined by
quantum correlation of the input state~\cite{Kitagawa,Wineland,Ma,Toth,Lucke}%
. For a twin-Fock state, the quasi-probability distribution spreads along
the equator of the Poincar\'{e} sphere; see Fig.~\ref{fig1}(b). This
observation suggests that the phase generator can take the form $G=J_{x}\cos
\alpha +J_{y}\sin \alpha $ for arbitrary $\alpha $ ($=\pi /2$ in Eq.~(\ref%
{Pn1n2})), which results in the quantum Fisher information $F_{Q}=N(N+2)/2$
and hence the quantum Cram\'{e}r-Rao bound $\delta \theta _{\mathrm{QCRB}}=1/%
\sqrt{F_{Q}}\simeq \sqrt{2}/N$. Therefore, the enhancement factor is given
by $\eta =1/(\sqrt{N}\delta \theta _{\mathrm{QCRB}})=\sqrt{(N+2)/2}$.

The sensitivity diverges at certain values of $\theta $ (e.g., $\theta =0$, $%
\pm \theta _{\mathrm{dark}}$). This is because at those points, the slope of
the signal $\partial \langle \mu (\theta )\rangle /\partial \theta =0$, but $%
\Delta \mu \neq 0$, so that $\delta \theta \rightarrow \infty $ (see also
Appendix B). Such a singularity could take place for any finite-$N$ input
state, e.g., a single-photon state $|1,0\rangle $ and the multi-photon N00N
states~\cite{Israel}. For a general binary-outcome measurement, we show that
the inversion estimator is indeed the same as the asymptotically optimal
maximum likelihood estimator (MLE)~\cite{Fisher}, so the same divergence
also occurs for the MLE (see Appendix A). This problem can not be completely
avoided even when all the $(N+1)$ outcomes are taken into account.

\section{Simulated microscopy images}

To reconstruct the microscopy images, one first calibrates the interferometer (with no
sample present, as done in Ref \cite{Israel}) to obtain the averaged signal $%
P_{\mathrm{fit}}(+|\theta )$ as a function of the phase shift $\theta$. Next, at each spatial point of the sample, one
performs the binary-outcome measurements for $\mathcal{N}$ times to record the
occurrence frequency for the detection event of
interest, and then inverts the averaged signal $P_{\mathrm{fit}}(+|\theta )=%
\mathcal{N}_{+}(x,y)/\mathcal{N}$ to obtain the inversion estimator $\theta
_{\mathrm{est}}(x,y)$. If an offset phase $\varphi $ is applied before the
sample, then the estimator becomes $\phi _{\mathrm{est}}(x,y)=\theta _{%
\mathrm{est}}(x,y)-\varphi $~\cite{Israel}, where the offset phase $\varphi $
is chosen such that the total phase shift $\theta =\varphi +\phi (x,y)\in
\lbrack \theta _{\min },\theta _{\mathrm{dark}})$~\cite{senrange}.

The birefringence phase shift used here is $\phi (x,y)=0.1+0.437\cos
^{6}[2(x-\pi /2)^{2}+y^{2}]\in (0.1,0.537]$, which can be discretized into
pixels $(i,j)$, with $i,j=0,1,2,\cdots $. At each pixel, performing the
photon-counting measurements for $\mathcal{N}$ times and inverting the
signal, one can obtain the inversion estimator $\phi _{\mathrm{est}%
}(i,j)=\theta _{\mathrm{est}}(i,j)-\varphi $, where $\theta _{\mathrm{est}}$
is a solution to $P_{\mathrm{fit}}(+|\theta )=\mathcal{N}_{+}(i,j)/\mathcal{N%
}$. For each input twin-Fock state, $P_{\mathrm{fit}}(+|\theta )$ has been obtained from the
calibration of the interferometer (see the blue solid lines of Fig.~\ref%
{fig1}(c), and also Appendix B), and $\mathcal{N}_{+}(i,j)$ denotes the
occurrence number of the outcome \textquotedblleft $+$" at the pixel $(i,j)$.

To simulate the microscopy illuminated by a classical light, we consider a
single-photon state $|1,0\rangle $ as the input and treat the detection
event $n_{1}=1$ and $n_{2}=0$ as the outcome \textquotedblleft $+$%
\textquotedblright , which occurs with probability $P(+|\theta )=\cos
^{2}(\theta /2)$. Photon counting over the other outcome gives $P(-|\theta
)=\sin ^{2}(\theta /2)$, as demonstrated recently by Israel \textit{et al}~%
\cite{Israel}. Both of them exhibit the same phase dependence as that of a
coherent-state input light $|\alpha \rangle \otimes |0\rangle $~\cite%
{Distante,XMFeng}.

\begin{figure}[hptb]
\begin{centering}
\includegraphics[width=1\columnwidth]{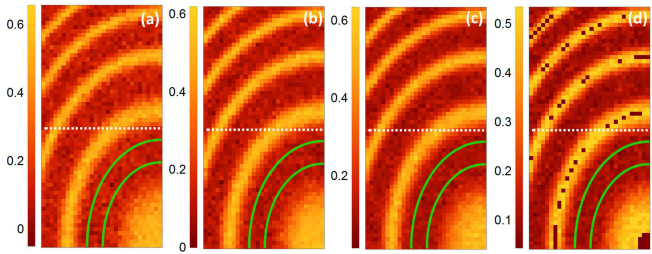}
\caption{Simulated microscopy images ($30\times 60$ pixels) reconstructed
from the phase estimator $\protect\phi_{\mathrm{est}}(i, j)$ for the
single-photon state (a), and the twin-Fock states with $N=2n=2$ (b), $4$
(c), and $6$ (d). The number of photons at each pixel $\mathcal{N}\times N=600$. Within the area enclosed by the green solid lines, the
phase shift sensed by the beams is almost optimal, and numerical simulation of the local standard deviation from $20$ repetitions gives $\mathrm{LSD}_{|1,0\rangle }=0.0413$, $\mathrm{LSD}_{|1,1\rangle }=0.0297$, $\mathrm{LSD}_{|2,2\rangle }=0.0253$, and $\mathrm{LSD}_{|3,3\rangle }=0.022$, indicating $\mathrm{LSD}_{|1,0\rangle }/\mathrm{LSD}_{|n,n\rangle}\approx \sqrt{(N+2)/2}$.  }
\label{fig2}
\end{centering}
\end{figure}

Figure~\ref{fig2} shows the simulated microscopy images using the inversion
estimator $\phi _{\mathrm{est}}(i,j)$ for the input twin-Fock states $%
|n,n\rangle $ with $n=N/2=1,2,3$, and that of the single-photon state $%
|1,0\rangle $. To keep exactly $600$ photons at each pixel, we use the
number of measurements $\mathcal{N}=600$ (a), $300$ (b), $150$ (c), and $100$
(d). From Fig.~\ref{fig2}(d), one can note that for the $6$-photon state $%
|3,3\rangle $, the simulated microscopy image is \emph{less} accurate at
some spatial points (see the speckles). This is because the sensed phase
shift $\theta =\varphi +\phi (i,j)\sim \theta _{\mathrm{dark}}$, at which
the phase sensitivity diverges. Similar phenomenon takes place for the
triphoton N00N state~\cite{Israel}, and also for any finite-$N$ photon state.

The image quality is improved with the quantum source of the light as long
as the sensed phase shift is far from the singular points~\cite{Israel}. To
quantify such a improvement, we calculate standard deviation of $\phi _{%
\mathrm{est}}(i,j)$ within a local area enclosed by the green solid lines of
Fig.~\ref{fig2}, as denoted by $\mathrm{LSD}_{|\psi _{\mathrm{in}}\rangle }$%
. Similar to Ref.~\cite{Israel}, we focus on the relative noise $\mathrm{LSD}%
_{|1,0\rangle }/\mathrm{LSD}_{|n,n\rangle }$, which gives a measure of the
improvement in the image quality beyond the classical illumination. From
each image of Fig.~\ref{fig2}, one can extract $\mathrm{LSD}_{|\psi _{%
\mathrm{in}}\rangle }$ and hence the relative noise. Taking $20$ pictures
for each input state, we obtain $\mathrm{LSD}_{|1,0\rangle }/\mathrm{LSD}%
_{|n,n\rangle }=1.39$ (for $n=N/2=1$), $1.63$ ($n=2$), and $1.88$ ($n=3$),
in agreement with the enhancement factor $\eta $.

\section{Phase locking to the optimal working point}

Due to the divergence of the phase sensitivity, the sensing range of the
quantum-enhance microscopy becomes narrow, especially when a higher-$N$
nonclassical state is injected. In order to remedy this problem, we propose
a scheme to control the offset phase at each spatial point of the sample
according to three estimators nearby, as illustrated schematically by Fig.~%
\ref{fig1}(a).

The basic idea is to insert a spatially dependent offset phase $\varphi (i,j)
$, such that the total phase sensed by the beams is close to the optimal
working point: $\theta (i,j)\equiv \varphi (i,j)+\phi (i,j)\sim \theta
_{\min }$. To determine the offset phase, we need some prior information to
the unknown phase $\phi (i,j)$ before the measurements. Quantum measurements
with adaptive feedback maximizes the information gain in subsequent
measurements and have been experimentally shown to be a powerful technique
to achieve the precision beyond the shot-noise limit~\cite{GYXiang10,Higgins}%
. However, application of the existing feedback-based phase estimation (see
e.g., Ref.~\cite{Berry2000}) in the microscopy is generally very
challenging. For our binary-outcome measurements, a global feedback strategy
for $\mathcal{N}_{\mathrm{tot}}$ measurements requires solving a set of
nonlinear equations with $2^{\mathcal{N}_{\mathrm{tot}}+1}-1$ unknown
variables~\cite{Berry2000}. Recently, Hentschel and Sanders~\cite{Sanders}
proposed an approximate scheme that reduces the number of unknown variables
to $\sim O(\mathcal{N}_{\mathrm{tot}})$. Here we are interested in
estimating the values of the phases at all the pixels of the sample, which
typically requires $\mathcal{N}_{\mathrm{tot}}=\mathcal{N}\times N_{\mathrm{%
pixels}}\sim 10^{6}$, where $N_{\mathrm{pixels}}$ denotes total number of
pixels. In this case, even the approximate strategy becomes formidable.

We present a simple but effective scheme that adjusts the offset phase after
every $\mathcal{N}$ measurements per pixel. Specially, we first estimate the
true value of phase shift at the pixel $(0,0)$, e.g., $\phi _{\mathrm{est}%
}(0,0)\simeq 0.1\mathrm{rad}$. From the starting point, we can obtain all
the estimators by adjusting the offset phase as illustrated in Fig.~\ref%
{fig3}(a) and (b). For instance, to estimate $\phi (1,0)$, we adjust the
offset phase as $\varphi (1,0)=$ $\theta _{\min }-\phi _{\mathrm{est}}(0,0)$%
, which ensures the phase locking to the optimal working point $\theta
(1,0)=\varphi (1,0)+\phi (1,0)\simeq \theta _{\min }$, provided $\phi
(1,0)\simeq \phi _{\mathrm{est}}(0,0)$. With this offset phase, one performs
$\mathcal{N}$ measurements at the pixel $(1,0)$ to obtain a local phase
estimator $\phi _{\mathrm{est}}(1,0)=\theta _{\mathrm{est}}-\varphi (1,0)$,
where $\theta _{\mathrm{est}}$ is a solution to $P_{\mathrm{fit}}(+|\theta )=%
\mathcal{N}_{+}(1,0)/\mathcal{N}$. Similarly, one can obtain the estimator $%
\phi _{\mathrm{est}}(0,1)$. To estimate $\phi (1,1)$, we use the three
estimators in a rectangle and adjust the offset phase to $\varphi
(1,1)=\theta _{\min }-[\phi _{\mathrm{est}}(0,1)+\phi _{\mathrm{est}%
}(0,0)+\phi _{\mathrm{est}}(1,0)]/3$, which helps to lock $\theta
(1,1)=\varphi (1,1)+\phi (1,1)$ at the pixel $(1,1)$ to the optimal working
point $\theta _{\min }$. Repeating the above procedures, one can measure the
phase of all the pixels over the entire sample.

\begin{figure}[hptb]
\begin{centering}
\includegraphics[width=1\columnwidth]{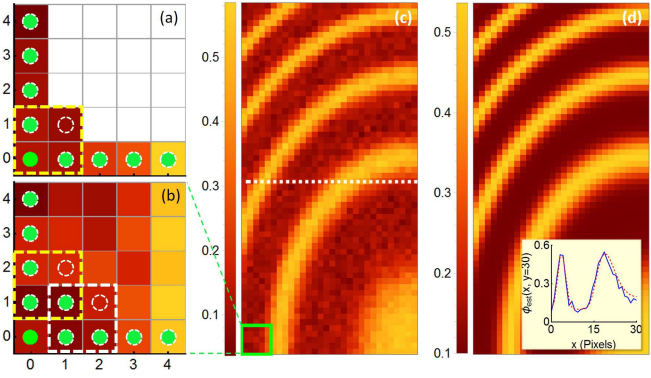}
\caption{Two steps of the phase locking (a) and (b), the simulated microscopy
image for the input 6-photon state $|3,3\rangle$ (c), and the true
value of phase shift (d). In (a), the offset phase is tuned as $\protect%
\varphi (i, 0)=\protect\theta_{\min}-\protect\phi_{\mathrm{est}}(i-1,0)$,
and $\protect\varphi (0,j)=\protect\theta_{\min}-\protect\phi_{\mathrm{est}%
}(0, j-1)$; in (b), it becomes $\protect\varphi (i,j)=\protect\theta_{\min}-[%
\protect\phi_{\mathrm{est}}(i-1,j)+\protect\phi_{\mathrm{est}}(i,j-1)+%
\protect\phi_{\mathrm{est}}(i-1,j-1)]/3$. In (c) $\mathcal{N}=100$ to
keep exactly $600$ photons at each pixel. The inset in (d): the simulated
estimators at the pixel $y=30$ as a function of $x$ (blue solid) and that of the true value of phase shift (red dashed).}
\label{fig3}
\end{centering}
\end{figure}

In Fig.~\ref{fig3}(c), we show the microscopy image for the 6-photon state $%
|3,3\rangle $ using the phase locking method. The main advantage of this
method is that the singular points (i.e., the speckles) disappear.
Furthermore, compared with previous adaptive feedback schemes~\cite%
{GYXiang10,Berry2000,Sanders} that adjusts a controllable phase after each
single measurement, our scheme updates the offset phase every $\mathcal{N}$
measurements. This costs much less computational resources, while it can
still improve the image quality significantly. The overall quality of the
image can be quantified by the root-mean-square error, i.e., $\mathrm{RMSE}=%
\sqrt{\sum_{i,j}[\phi _{\mathrm{est}}(i,j)-\phi (i,j)]^{2}/N_{\mathrm{pixels}%
}}$, which approaches the optimal value of the standard deviation $\mathrm{%
LSD}_{|3,3\rangle }=0.022$, as depicted in Fig.~\ref{fig2}(d). This
observation implies that at most of the pixels, the phase shift sensed by
the beams is optimal.

\section{Combination of two binary-outcome measurements}

The phase locking scheme requires control of the feedback phase after every $%
\mathcal{N}$ measurements at each pixel. To further reduce the cost, one can
use a fixed offset phase $\varphi $ (as implemented experimentally in Ref.~%
\cite{Israel}) and then perform two sequences of binary-outcome photon
counting measurements: one sequence with the offset phase $\varphi $ and the
other sequence without any offset phase. Then we combine
all the measurement results to obtain the MLE and hence the microscopy images, i.e., $\phi
_{\mathrm{mle}}(i,j)$.

\begin{figure}[hptb]
\begin{centering}
\includegraphics[width=1\columnwidth]{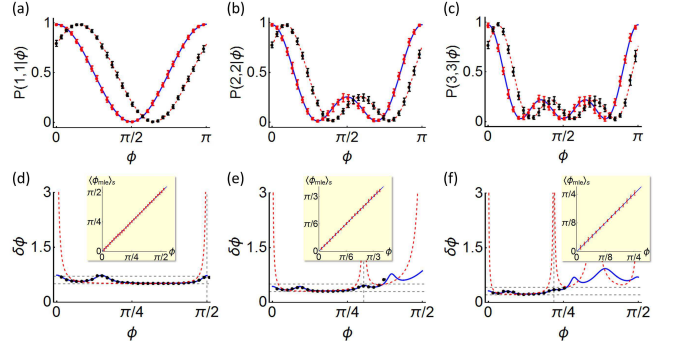}
\caption{Simulated count rates (a)-(c) and uncertainty of the MLE (d)-(f) for $\varphi =-0.3\times \protect\theta _{\mathrm{dark}}$ and $\mathcal{N}_{1}=\mathcal{N}_{2}=\mathcal{N}%
/2$, where $N\times\mathcal{N}=1200$ is fixed for the input states $|1,1\rangle $ (left), $|2,2\rangle $ (middle), and $|3,3\rangle $ (right). Red dashed lines in (d)-(f): the phase sensitivity $1/\sqrt{F(\phi)}$, with the locations of $\theta _{\mathrm{dark}}$ indicated by the vertical lines; Blue solid lines: the sensitivity with the total Fisher information (see text). The inset: statistical average of $\protect\phi_{\mathrm{mle}}$ as a function of $\protect\phi$ for $20$ repetitions.}
\label{fig4}
\end{centering}
\end{figure}

Following Ref.~\cite{Israel}, let us begin with the calibration the
interferometer using different known values of phase shift $\phi $ and a
fixed offset phase $\varphi $ for each input state. Performing $\mathcal{N}%
_{1}$ measurements without the offset phase, one can obtain the occurrence
number $\mathcal{N}_{1}^{(+)}$ for the outcome $n_{1}=n_{2}=n$. In the
presence of the offset phase, one performs another $\mathcal{N}_{2}$
measurements over the output state $\exp [-i(\phi +\varphi
)J_{y}]|n,n\rangle $ to obtain the occurrence number $\mathcal{N}_{2}^{(+)}$%
. In the upper panel of Fig.~\ref{fig4}, we plot the averaged count rates $%
\mathcal{N}_{1}^{(+)}/\mathcal{N}_{1}$ and $\mathcal{N}_{2}^{(+)}/\mathcal{N}%
_{2}$ (the circles) as functions of $\phi $ and fit them as $P_{\mathrm{fit}%
}(+|\phi )$ (the blue solid) and $P_{\mathrm{fit}}(+|\varphi +\phi )$ (the
red dashed), respectively.

Next, we perform the above binary-outcome photon counting at each pixel of
the sample for totally $\mathcal{N}$ ($=\mathcal{N}_{1}+\mathcal{N}_{2}$)
measurements to retrieve $\phi _{\mathrm{mle}}$ that maximizes the
likelihood function:
\begin{eqnarray}
\mathcal{L}(\phi )\!\! &\varpropto &\!\!\left[ P_{\mathrm{fit}}\left(
+|\varphi +\phi \right) \right] ^{\mathcal{N}_{2}^{(+)}}\left[ 1-P_{\mathrm{%
fit}}\left( +|\varphi +\phi \right) \right] ^{\mathcal{N}_{2}-\mathcal{N}%
_{2}^{(+)}}  \notag \\
&&\times \left[ P_{\mathrm{fit}}\left( +|\phi \right) \right] ^{\mathcal{N}%
_{1}^{(+)}}\left[ 1-P_{\mathrm{fit}}\left( +|\phi \right) \right] ^{\mathcal{%
N}_{1}-\mathcal{N}_{1}^{(+)}},  \label{likelihood2}
\end{eqnarray}%
where the occurrence numbers $\mathcal{N}_{1}^{(+)}$ and $\mathcal{N}%
_{2}^{(+)}$ are spatially dependent, containing phase information of the
sample. At each pixel $(i,j)$, the phase estimator $\phi _{\mathrm{mle}}$
and its uncertainty $\sigma $ can be obtained by numerically finding the
peak of the likelihood function and the  $68.3\%$ confidence interval around
the peak~\cite{Pezze&Smerzi}. The inset of Figs.~\ref{fig4}(d)-(f) shows
statistical average of the estimator $\langle \phi _{\mathrm{mle}}\rangle
_{s}=\phi $, indicating that $\phi _{\mathrm{mle}}$ is unbiased for $\phi
\in (0,\theta _{\mathrm{dark}})$. Interestingly, we find that the averaged
phase uncertainty per measurement $\sqrt{\mathcal{N}}\langle \sigma \rangle
_{s}$ (the circles) follows the lower bound of the phase sensitivity $\delta
\phi \!=\!\sqrt{\mathcal{N}}/\sqrt{F_{\mathrm{tot}}(\phi )}$ (the blue solid
lines), where $F_{\mathrm{tot}}(\phi )=\mathcal{N}_{1}F(\phi )+\mathcal{N}%
_{2}F(\phi +\varphi )$ is the total Fisher information of all the two
sequences of binary-outcome measurements and $F(\phi)$ is the Fisher information of a single sequence of the measurements (see Appendix A, Eq.~(\ref{CFI})).
Obviously, the singularity of $\delta \phi $ can be completely eliminated by
a suitable choice of the offset phase $\varphi $ (which maximizes the total
Fisher information), in sharp contrast to the previous result~\cite%
{Israel,senrange}.

In Fig.~\ref{fig4}, we show that with a fixed offset phase $\varphi
=-0.3\times \theta _{\mathrm{dark}}$ for each input twin-Fock state, the
unbiased estimator $\phi _{\mathrm{mle}}$ does not show any singularity and
its uncertainty can surpass the shot-noise limit as $\phi $ increases up to $%
\sim \theta _{\mathrm{dark}}$. It is therefore useful for estimating the
phase information of a sample at the sub-shot-noise limit through $\phi _{%
\mathrm{mle}}(i,j)$.

\section{Conclusion}

In summary, we have investigated theoretically the binary-outcome photon
counting and its potential applications in quantum-enhanced microscopy using
the input twin-Fock states of light. Our results show that the inversion
estimator is the same to the asymptotically optimal maximum likelihood
estimator. Both estimators may suffer from a divergent uncertainty that
reduces the quality of the microscopy images. To remedy this problem, we
propose a simple method to lock the phase shift sensed by the beams at the
optimal working point with a spatially dependent offset phase. The overall
image quality outperforms the case of classical light illumination by a
factor $\sim \sqrt{(N+2)/2}$. We further show that a combination of two
sequences of binary-outcome photon counting measurements, one sequence with
a fixed offset phase and the other sequence without any offset phase, also
works to remove the singularity. Our results remain valid for any kind of
binary-outcome measurement and pave the way for realistic implementations of
quantum-enhanced microscopy that uses high-$N$ nonclassical states of the
light.

\begin{acknowledgments}
We would like to thank S. Rosen and Y. Silberberg for kindly response to our
questions, as well as P. Liu and T. Li for their assistance in plotting the
figures. This work has been supported by the NSFC (Grant Nos. 11421063,
11534002, 11274036, and 11322542), the National 973 program (Grant Nos.
2012CB922104, 2014CB921403, and 2014CB848700). G.R.J. also acknowledges
support from the Major Research Plan of the NSFC (Grant No. 91636108).
\end{acknowledgments}

\appendix
\renewcommand{\labelenumi}{(\roman{enumi})}

\section{Phase estimators for a general binary-outcome measurement}

In the following, we introduce the concept of binary-outcome measurement and
present the details of our numerical simulations.

Binary-outcome measurements have been widely adopted in quantum metrology~%
\cite{Bollinger,Dowling,Gerry,Cohen,Brivio,Distante,XMFeng}, and recently in
quantum-enhanced microscopy~\cite{Ono,Israel}. As the simplest measurement
scheme, the output signal can be expressed as
\begin{equation}
\langle \mu (\theta )\rangle =\sum_{i=\pm }\mu _{i}P\left( i|\theta \right)
\approx \sum_{i=\pm }\mu _{i}\frac{\mathcal{N}_{i}}{\mathcal{N}},
\label{signal}
\end{equation}%
where $\mathcal{N}_{\pm }/\mathcal{N}$ denotes the occurrence frequency of
the outcome $\mu _{\pm }$, measured by the normalized coincidence rate with
a finite number of photon counts $\mathcal{N}=\mathcal{N}_{+}+\mathcal{N}%
_{-} $. For the input twin-Fock states $|n, n\rangle$~\cite%
{FWSun,GYXiang10,Xiang}, the specific detection event $n_{1}=n_{2}=n$ is of
interest and can be treated as the outcome \textquotedblleft $+$" and the
others as \textquotedblleft $-$", with the conditional probability $%
P(+|\theta)\equiv P(n,n|\theta)$ and hence $P(-|\theta)=1-P(+|\theta)$.
Taking $\mu _{+}=+1$ and $\mu _{-}=0$, the signal becomes $\langle \mu
(\theta )\rangle =P( +|\theta )=P(n, n|\theta)$, as expected. Similarly, the
parity detection gives two outcome $\pm 1$, according to even or odd number
of photons being detected at one port of the interferometer~\cite%
{Bollinger,Dowling,Gerry,Cohen}. Recently, quantum-enhanced microscopy with
a two-photon N00N state has been demonstrated by counting odd number of
photons~\cite{Ono}. For a measurement with continuous-variable outcome, one
can also realize a binary-outcome measurement by dividing the date into two
bins~\cite{Distante}. These cases are indeed binary-outcome measurement~\cite%
{XMFeng}.

For any kind of binary-outcome measurement, the inversion estimator $\theta
_{\mathrm{est}}$ can be obtained by inverting the averaged signal, which is
indeed a solution of Eq.~(\ref{signal}), or equivalently $P(+|\theta )=%
\mathcal{N}_{+}/\mathcal{N}$, independently from the measured values $\mu
_{\pm }$. According to the error propagation, the uncertainty of $\theta _{%
\mathrm{est}}$ depends on the fluctuations of signal $\Delta \mu =(\mu
_{+}-\mu _{-})\Delta \mathcal{N}_{+}/\mathcal{N}$, with $\Delta \mathcal{N}%
_{+}=\sqrt{\mathcal{N}P(+|\theta )P(-|\theta )}$ being standard deviation of
a binomial distribution:
\begin{equation}
\mathcal{L}(\theta ;\mathcal{N}_{+})=\binom{\mathcal{N}}{\mathcal{N}_{+}}%
\left[ P\left( +|\theta \right) \right] ^{\mathcal{N}_{+}}\left[ P\left(
-|\theta \right) \right] ^{\mathcal{N}_{-}},  \label{binomial}
\end{equation}%
where $\binom{n}{k}$ is the binomial coefficient, $P(+|\theta )+P(-|\theta
)=1$, and hence $\sum_{\mathcal{N}_{+}}\mathcal{L}(\theta ;\mathcal{N}%
_{+})=[P(+|\theta )+P(-|\theta )]^{\mathcal{N}}=1$. On the other hand, from
Eq.~(\ref{signal}), we obtain the slope of signal $\partial \langle\mu
(\theta )\rangle/\partial \theta =(\mu _{+}-\mu _{-})\partial P(+|\theta
)/\partial \theta $, which, together with $\Delta \mu $, gives the phase
uncertainty
\begin{equation}
\delta \theta =\frac{\Delta \mu }{\left\vert \partial \langle\mu (\theta
)\rangle/\partial \theta \right\vert }=\frac{\sqrt{P\left( +|\theta \right)
P\left( -|\theta \right) }}{\sqrt{\mathcal{N}}\left\vert \partial P\left(
+|\theta \right) /\partial \theta \right\vert }=\frac{1}{\sqrt{\mathcal{N}%
F(\theta )}},  \label{uncertainty}
\end{equation}%
where, for a single-shot measurement, the classical Fisher information is
given by%
\begin{equation}
F(\theta )=\sum_{i=\pm }\frac{1}{P\left( i|\theta \right) }\left[ \frac{%
\partial P\left( i|\theta \right) }{\partial \theta }\right] ^{2}.
\label{CFI}
\end{equation}%
Our above results indicate that for any binary-outcome measurements with $%
\mathcal{N}\gg 1$, the simplest data processing based on inverting the
averaged signal always saturates the Cram\'{e}r-Rao lower bound~\cite{XMFeng}. This is somewhat counter intuitive since, according
to Fisher's theorem~\cite{Fisher}, this bound is saturable by maximum
likelihood estimator (MLE) as the number of measurements $\mathcal{N}\gg 1$.
To understand it, we further investigate the MLE by finding a value of $%
\theta $ that maximizes Eq.~(\ref{binomial}); Hereinafter, denoted by $%
\theta _{\mathrm{mle}}$. When $\mathcal{N}_{\pm }\sim O(\mathcal{N})\gg 1$,
the binomial distribution of $\mathcal{L}(\theta ;\mathcal{N}_{+})$ becomes
normal
\begin{equation}
\mathcal{L}(\theta ;\mathcal{N}_{+})\varpropto \exp \left( -\frac{\left[
\mathcal{N}_{+}-\mathcal{N}P\left( +|\theta \right) \right] ^{2}}{2(\Delta
\mathcal{N}_{+})^{2}}\right) ,  \label{Guassian}
\end{equation}%
which indicates that the MLE $\theta _{\mathrm{mle}}$ also satisfy the
equation $P(+|\theta )=\mathcal{N}_{+}/\mathcal{N}$, the same to that of $%
\theta _{\mathrm{est}}$.

The phase estimator $\theta _{\mathrm{mle}}$ and its uncertainty can be
obtained by maximizing Eq.~(\ref{binomial}). To avoid the phase ambiguity~%
\cite{Pezze,Higgins,Berry}, we introduce prior knowledge about the true
value of $\theta $ by assuming the prior probability $P(\theta )=1$ for $%
\theta \in (0,\theta _{\mathrm{dark}})$, and $0 $ outside, where $\theta _{%
\mathrm{dark}}$ denotes the location of the first dark fringe; see the
vertical dashed lines in Fig.~\ref{fig1}(c). Next, we fit the phase
distribution as a Gaussian around its peak~\cite{Pezze&Smerzi}, i.e.,
\begin{equation*}
\mathcal{P}(\theta |\mathcal{N}_{+})=CP(\theta )\mathcal{L}(\theta ;\mathcal{%
N}_{+})\propto\exp \left[ -\frac{(\theta -\theta _{\mathrm{mle}})^{2}}{%
2\sigma ^{2}}\right] ,  \label{phase distri}
\end{equation*}%
where $C$ is a normalized factor, and $\sigma $ is $68.3\%$ confidence
interval of the Gaussian around $\theta _{\mathrm{mle}}$, given by
\begin{equation}
\sigma \simeq \sqrt{\frac{C}{|\partial ^{2}\mathcal{P}(\theta |\mathcal{N}%
_{+})/\partial \theta ^{2}|_{\theta=\theta _{\mathrm{mle}}}}}.  \label{cig}
\end{equation}
The above results remain valid for any input state of the probes and is
independent from any specific form of the noise. For the input twin-Fock
states, the averaged phase uncertainty of the MLE, i.e., $\sqrt{\mathcal{N}}%
\langle \sigma \rangle _{s}$ (the circles of Fig.~\ref{fig1}(d)), shows a
good agreement with the sensitivity per measurement $1/\sqrt{F(\theta)}$
(the blue solid line), where $\langle (...)\rangle _{s}\equiv
\sum_{i=1}^{M}(...)_{i}/M$ denotes the statistical average for $M$
repetition of measurements.

\section{Numerical simulations}

We consider a single-photon state $|1,0\rangle $ as the input to simulate
the microscopy with a classical illumination~\cite{Ono,Israel}. It is easy
to obtain the conditional probability for detecting a single photon in the
horizontal polarization mode and vacuum in the vertical polarization mode,
i.e., $P(1,0|\theta )=|\langle 1,0|\exp (-i\theta J_{y})|1,0\rangle
|^{2}=\cos ^{2}(\theta /2)$. If we treat the detection event $n_{1}=1$ and $%
n_{2}=0$ as the outcome \textquotedblleft $+$", and the others as
\textquotedblleft $-$", then this is indeed a binary-outcome photon counting
measurement, with the output signal $\langle \mu (\theta )\rangle
=P(+|\theta )=\cos ^{2}(\theta /2)$. From Eq.~(\ref{uncertainty}), we
immediately obtain the phase sensitivity $\delta \theta =1/\sqrt{\mathcal{N}%
F(\theta )}$, where the classical Fisher information is given by%
\begin{equation}
F(\theta )=\frac{1}{P\left(+|\theta \right) [1-P\left(+|\theta \right) ]}%
\left[ \frac{\partial P\left(+|\theta \right) }{\partial \theta }\right]
^{2}=1,  \label{Fisher10}
\end{equation}%
which is independent from the true value of phase shift $\theta $.

In real experiment, e.g., Ref.~\cite{Israel}, the achievable sensitivity
depends on $\theta $, arising from the detection efficiency, the photon
loss, the imperfect visibility, and so on. To take the experimental
imperfections into account, we first rewrite Eq. (1) in the main text as
\begin{equation}
P\left( n_{1},n_{2}|\theta \right) \rightarrow \frac{2hV}{1+V}P\left(
n_{1},n_{2}|\theta \right) +\frac{h(1-V)}{1+V},  \label{Imperfect}
\end{equation}%
where the peak height $h$ and the visibility $V$, as shown in Table~\ref{Tab}%
, can be determined by the photon-counting measurement. Next, we randomly
choose $\mathcal{N}$ values of the outcomes according to $%
P(n_{1},n_{2}|\theta )$ for each a given $\theta $~\cite{Pezze&Smerzi}.
Specially, for the input $|1,0\rangle $, we generate $\mathcal{N}$ random
numbers $\{\xi _{1},\xi _{2},...,\xi _{\mathcal{N}}\}$, where $\xi _{k}\in
\lbrack 0,1]$ for $k=1$, $2$, ..., $\mathcal{N}$. If $0\leq \xi
_{k}<P(1,0|\theta )$, we set $\xi _{k}=+1$, otherwise, $\xi _{k}=0$, then
the number of \textquotedblleft $+1$" can be used to simulate the occurrence
number of the event $n_{1}=1$ and $n_{2}=0$, denoted as $\mathcal{N}_{+}$.
Finally, for each a given $\theta \in (-\pi ,\pi )$, we repeat the above
simulations for $M$ times to obtain the averaged signal $\langle \mathcal{N}%
_{+}\rangle _{s}/\mathcal{N}$ and fit it as $P_{\mathrm{fit}}(1,0|\theta )$.

\begin{table}[htbp]
\caption{For the single-photon state $|1, 0\rangle$ and the twin-Fock states
$|n, n\rangle$ with $n=N/2=1$, $2$, and $3$, the parameters used in the
simulations.}
\label{Tab}\vspace{0.1cm}
\begin{tabular}{ccccccccc}
\hline\hline
$N$ &  & 1 &  & 2 &  & 4 &  & 6 \\ \hline
$V$, $h$ &  & 0.994, 0.99 &  & 0.983, 0.985 &  & 0.97, 0.98 &  & 0.94, 0.975
\\[0.1cm] \hline\hline
\end{tabular}
\vspace{0.1cm}
\end{table}

\begin{figure}[hbpt]
\begin{centering}
\includegraphics[width=1\columnwidth]{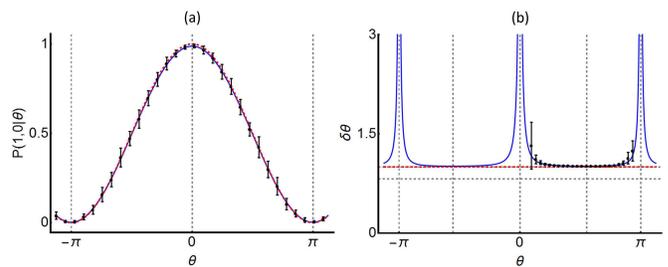}
\caption{Statistical average of $\mathcal{N}_{+}/\mathcal{N}$ (a) and $\protect%
\sqrt{\mathcal{N}}\protect\sigma$ (b) for the single-photon input state, with the number of photon counts $\mathcal{N}=100$ and the number of repetitions $%
M=20$, where $\protect\sigma$ is given by Eq.~(\ref{cig}). Red
dashed and blue solid lines: $P(1, 0|\protect\theta)$ and $P_{\mathrm{fit}%
}(1, 0|\protect\theta)$, and the associated sensitivities $1/\sqrt{F(\theta)}$. Vertical lines:
locations of $\theta=0$, $\pm\protect\theta_{\mathrm{dark}}$, and $\pm\protect\theta_{\min}$. The horizontal lines in (b): the shot-noise limit $1/%
\protect\sqrt{N}$ and the theoretical bound $\protect\sqrt{2}/\protect\sqrt{N(N+2)}$ for $N=1$.}
\label{figS1}
\end{centering}
\end{figure}

In Fig.~\ref{figS1}, we numerically simulate the binary-outcome photon
counting for the input state $|1,0\rangle $, using the parameters in Table~%
\ref{Tab}. For $\mathcal{N}=100$ and $M=20$, we obtain $P_{\mathrm{fit}%
}(1,0|\theta )=aP(1,0|\theta )+b$, with $a=0.988$ and $b=0.00396$.
Substituting it into the first result of Eq.~(\ref{Fisher10}), we further
obtain the phase sensitivity per measurement $\sqrt{\mathcal{N}}%
\delta\theta=1/\sqrt{F(\theta)}$; see the blue solid line. The optimal
working point for phase sensing is $\theta _{\min}=1.7371\sim \pi /2$ and
the best sensitivity $1/\sqrt{F(\theta _{\min })}=1.0116\sim 1$, as
predicted by Eq.~(\ref{Fisher10}). Our results coincide quite well with the
experimental data of Ref.~\cite{Israel}, where the signal $P(0,1|\theta
)=\sin ^{2}(\theta /2)$ was measured. Using $P_{\mathrm{fit}}(1,0|\theta)$,
we also calculate the phase uncertainty of the MLE, i.e., $\sqrt{\mathcal{N}}%
\langle \sigma \rangle _{s}$ (the circles), which shows a good agreement
with the sensitivity (the blue solid line).

To simulate the twin-Fock experiments~\cite{FWSun,GYXiang10,Xiang}, we first
write down exact results of the signal for the input states $|1,1\rangle $, $%
|2,2\rangle $, and $|3,3\rangle $, given by $P(1,1|\theta )=\cos ^{2}(\theta
)$, $P(2,2|\theta )=[1+3\cos (2\theta )]^{2}/16$~\cite{FWSun,GYXiang10}, and
$P(3,3|\theta )=[3\cos (\theta )+5\cos (3\theta )]^{2}/64$~\cite{Xiang},
respectively. Next, we generate $\mathcal{N}$ random numbers according to
Eq.~(\ref{Imperfect}) with the parameters in Table~\ref{Tab}. The averaged
signal and the associated phase sensitivity are shown in Fig.~\ref{fig1}(c)
and (d).

Note that the exact result of $P(+|\theta )$ and the choice of random
numbers using Eq.~(\ref{Imperfect}) are unneccessary as long as the counts rate has been recorded in real experiment. Furthermore, the phase
sensitivity diverges at certain values of $\theta $. Formally, this is
because the slope of signal $\partial P(+|\theta )/\partial \theta =0$, but
the variance of signal $(\Delta \mu )^{2}\propto P(+|\theta )[1-P(+|\theta
)]\neq 0$. Here, the outcome \textquotedblleft $+$" represents $n_{1}=1$ and
$n_{2}=0$ for the input state $|1,0\rangle $; While for the twin-Fock states
$|n,n\rangle $, it stands for the detection event $n_{1}=n_{2}=n$. Due to
the experimental imperfections, the signal $P_{\mathrm{fit}}(+|\theta )\neq
0,1$ at certain values of phase shift (e.g., $\theta =0$, $\pm \theta _{%
\mathrm{dark}}$), so the variance of signal is nonvanishing at that points
but the slope of signal is still vanishing, which leads to the singularity
of the sensitivity.


\begin{thebibliography}{99}
\bibitem{senrange} Using the method of Ref.[5], the sensing range of a
quantum-enhanced microscopy is about $|\theta_{\mathrm{dark}}-\theta_{\min}|$%
, limited by the optimal work point for phase sensing $\theta_{\min}$ and
the location of dark fringe nearby, $\theta_{\mathrm{dark}}$.
\end{thebibliography}
\end{document}